# Deep photonic reservoir computing recurrent network

CHENG WANG[1,2*]

[1]*School of Information Science and Technology, ShanghaiTech University, Shanghai 201210, China*
[2]*Shanghai Engineering Research Center of Energy Efficient and Custom AI IC, ShanghaiTech University, Shanghai 201210, China*
*\*Corresponding author: wangcheng1@shanghaitech.edu.cn*

**Deep neural networks usually process information through multiple hidden layers. However, most hardware reservoir computing recurrent networks only have one hidden reservoir layer, which significantly limits the capability of solving real-world complex tasks. Here we show a deep photonic reservoir computing (PRC) architecture, which is constructed by cascading injection-locked semiconductor lasers. In particular, the connection between successive hidden layers is all optical, without any optical-electrical conversion or analog-digital conversion. The proof of concept is demonstrated on a PRC consisting of 4 hidden layers and 320 interconnected neurons. In addition, we apply the deep PRC in the real-world signal equalization of an optical fiber communication system. It is found that the deep PRC owns strong ability to compensate the nonlinearity of fibers.**

## 1. INTRODUCTION

Deep neural networks with multiple hidden layers have been substantially advancing the development of artificial intelligence. In comparison with the digital electronic computing based on the von Neumann architecture, optical computing can boost the energy efficiency while reduce the computation latency [1-3]. In recent years, a large variety of optical computing architectures have been proposed, and most focused on the linear multiply-accumulation operation [4-8]. Together with the nonlinear activation function in the digital domain, optical convolutional neural networks and multilayer perceptrons have been extensively demonstrated. In contrast to the above two feedforward neural networks, recurrent neural networks (RNN) have inherent memory effect and are favorable for solving time-dependent tasks such as natural language processing and temporal signal processing [9]. Reservoir computing (RC) is such a kind of RNN, but with fixed weights in the input layer and in the hidden reservoir layers [10,11]. Only weights in the readout layer require training, which leads to a simple training algorithm and a fast training speed. Optoelectronics-based [12-14] and memristor-based [15-17] RCs have been intensively investigated, while various types of hardware RCs have been discussed as well [18]. However, most hardware RCs only have one hidden reservoir layer, which substantially limits the capability of dealing with real-world problems. A comprehensive theoretical analysis by C. Gallicchio *et al.* has pointed out that the deep hierarchy of RCs owned multiple time scales and frequency components, and thereby boosted the richness of dynamics and the diversity of representations [19,20]. Several paradigms of combining multiple reservoirs have been theoretically compared in literatures, and it was found that a unidirectional coupling scheme of hidden reservoirs was beneficial to improve the performance of RCs [21,22]. Indeed, the deep configuration raises both the linear and the nonlinear memory capacities of RCs [23,24]. Interestingly, Penkovsky *et al.* showed that a deep RC with time-delay loops was equivalent to a deep convolutional neural network [25]. In experiment, Nakajima constructed a deep RC based on a Mach-Zehnder modulator associated with an optoelectronic feedback loop [26]. However, there is only one piece of hardware, which is reused in each hidden layer. The interconnection between successive layers requires optical-electrical conversion (OEC), analog-digital conversion (ADC), as well as the inverse conversions. The above four conversion processes consume high power and introduce a large amount of latency, which significantly counteract the merits of optical computing. Lupo *et al.* recently proposed a two-layer RC based on two groups of frequency combs, which were produced by the phase modulation of light [27]. The interconnection between the two layers is implemented in the electrical domain through the OEC. Nevertheless, the scalability of the RC depth is limited by its tradeoff with the width.

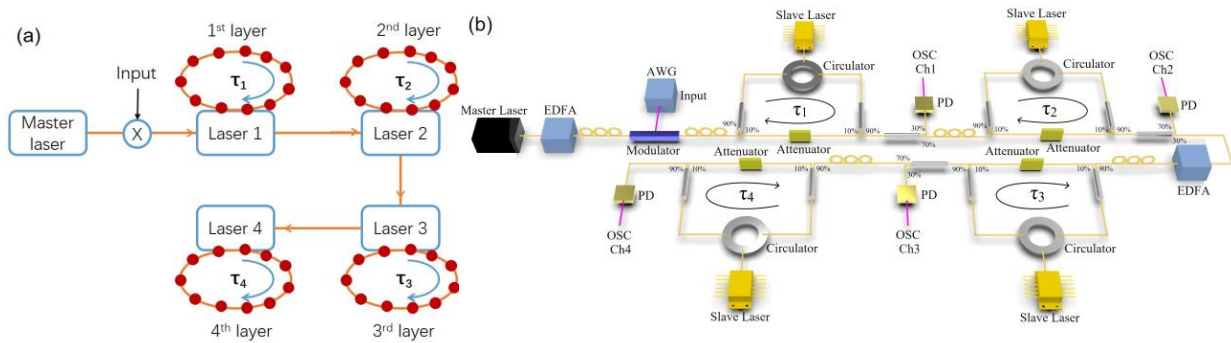

**Fig. 1.** (a) Schematic architecture of the deep PRC. (b) Experimental setup of the deep PRC. AWG: arbitrary waveform generator; OSC: oscilloscope; PD: photodiode; EDFA: erbium-dope fiber amplifier; Ch: channel. The hidden layers are interconnected by the optical injection. The optical feedback loops provides virtual neurons.

This work presents a deep PRC based on cascading injection-locked semiconductor lasers. The hidden-layer interconnections are fully optical without any OEC and ADC. The deep PRC architecture with 4 hidden layers and 320 neurons is successfully demonstrated in experiment. In particular, the PRC depth is highly scalable without any power and coherence limitation. The deep PRC is applied in the signal equalization of an optical fiber communication system. It is proved that the deep PRC has strong ability to mitigate the Kerr nonlinearity of optical fibers, and hence to improve the signal quality at the optical receiver.

## 2. DEEP PRC ARCHITECTURE AND EXPERIMENTAL SETUP

Figure 1(a) illustrates the architecture of the deep PRC. A single-mode master laser uni-directionally injects into the slave laser (Laser 1) in the first hidden layer of the reservoir. The optical injection is operated in the stable regime, which is bounded by the Hopf bifurcation and the saddle-node bifurcation [28,29]. Partial light of Laser 1 goes to the second layer of the reservoir and locks Laser 2 through optical injection. In the same way, Laser 2 locks Laser 3 in the third layer, and then Laser 3 locks Laser 4 in the fourth layer. As a result, the lasing frequencies of all the four slave lasers are locked to be the same as that of the master laser. Besides, the phases of all the slave lasers are synchronized with the master laser as well. In each hidden layer, the laser is subject to an optical feedback loop, which produces a large number of virtual neurons through nonlinear laser dynamics [14,30]. The optical feedback is also operated in the stable regime, which is separated from the unstable regime through a critical feedback level [28,31]. In the input layer, the input signal is multiplied by a random mask, and this pre-processed signal is superimposed onto the carrier wave of the master laser through an optical modulator. The masking process plays a crucial role in the PRC system. On one hand, the fast varying mask sequence maintains the instantaneous state of all the time-delay reservoirs [30,32]. On the other hand, the mask interval defines the interval of virtual neurons. The neuron number in each hidden layer is determined by the clock cycle divided by the neuron interval.

In the readout layer, the neuron states in all the four hidden layers are tracked simultaneously. The target value is obtained through the weighted sum of all the neuron states, and the weights are trained through the algorithm of ridge regression [32]. Based on the deep PRC scheme, Fig. 1(b) shows the corresponding experimental setup. A tunable external cavity laser (Santec TSL-710) serves as the master laser, and its output power is amplified by an erbium-dope fiber amplifier (EDFA). The polarization of the light is aligned with a Mach-Zehnder intensity modulator (EOSPACE, 40 GHz bandwidth) through a polarization controller. The input signal is multiplied by a random binary mask consisting of {1, 0}. This pre-processed signal is generated from an arbitrary waveform generator (AWG, Keysight 8195A, 25 GHz bandwidth), which then drives the modulator. The polarization of the modulated light is re-aligned with the polarization of the slave laser in the first hidden layer. The four slave lasers in the hidden layers are commercial Fabry-Perot lasers with multiple longitudinal modes. In each layer, the optical feedback loop is formed by an optical circulator and two 90:10 couplers. The feedback strength is adjusted by an optical attenuator. At the output of each hidden layer (except the fourth layer), 70% light is uni-directionally injected into the subsequent layer to lock the slave laser, and the polarization of the light is re-aligned. Between the second and the third layers, the laser power is amplified by using another EDFA. The neuron states of all the four layers are detected by broadband photodiodes (PD), and then recorded on the four channels (Ch) of a high-speed digital oscilloscope (OSC, Keysight DSAZ594A, 59 GHz bandwidth), simultaneously. The optical spectrum is measured by an optical spectrum analyzer with a resolution of 0.02 nm (Yokogawa). In the experiment, the time interval of neurons in each hidden layer is fixed at θ=0.05 ns, which is determined by the modulation rate of the optical modulator at 20 Gbps. The number of neurons in each layer is set at N=80, resulting in a total neuron number of 320 in the deep PRC of four hidden layers. Consequently, the clock cycle of the PRC system is $T_c$ =4.0 ns ($T_c$=θ×N). The sampling rate of the AWG is 60 GSa/s and the rate of the OSC is 80 GSa/s, respectively.

## 3. EXPERIMENTAL RESULTS

In the experiment, all the four FP lasers in the hidden layers exhibit an identical lasing threshold of Ith=8.0 mA. The pump currents and the corresponding output power of all the lasers are listed in Table 1, respectively. The delay times of the four optical feedback loops are fixed in the range of 63 to 68.5 ns, without any optimization. It is stressed that the delay times are more than 15 times longer than the clock cycle of the computing system, unlike the common synchronous case. Our recent work has proved that this asynchronous architecture is helpful to improve the PRC performance [29,33], owing to the rich neuron interconnections [34]. The feedback ratio is defined as the power ratio of the reflected light to the emitted light, which is set around -30 dB for all the four layers. The critical feedback level of the lasers is about -19.3 dB, and hence the optical feedback is operated in the stable regime. The injection ratio is defined as the ratio of the injected power from the laser in the previous layer to the emission power of the laser in the subsequent layer. As shown in Table 1, the injection ratios of each layer vary from about 2.0 up to 4.0. In addition, the detuning frequency is defined as the lasing frequency difference between the two lasers. All the detuning frequencies in Table 1 are set within the stable locking regime without optimization. Figure 1 shows the optical spectra of the FP lasers of multiple longitudinal modes in all the four reservoir layers. The spectrum peaks of the lasers are around 1550.98, 1542.63, 1548.91, and 1540.86 nm, respectively. Meanwhile, the free spectral ranges are 154.6, 154.8, 172.7, and 171.9 GHz, respectively. When applying optical injection from the master laser at 1546.5 nm, only one mode of the slave lasers closest to the injection wavelength is locked in the stable regime. All side modes are suppressed and the suppression ratio is more than 50 dB. This is because the optical injection reduces the gain of the laser medium [35].

**Table 1. Operation conditions of the deep PRC**

| Parameters | Layer 1 | Layer 2 | Layer 3 | Layer 4 |
|---|---|---|---|---|
| Laser current | 6.3×$I_{th}$ | 2.8×$I_{th}$ | 6.5×$I_{th}$ | 2.5×$I_{th}$ |
| Laser power | 9.0 mW | 2.1 mW | 11.2 mW | 2.4 mW |
| Feedback delay | 63.0 ns | 68.4 ns | 63.7 ns | 68.0 ns |
| Feedback ratio | -30.2 dB | -30.2 dB | -29.7 dB | -30.8 dB |
| Injection ratio | 4.0 | 2.2 | 3.8 | 1.9 |
| Detuning freq. | -55.0 GHz | -15.0 GHz | -33.7 GHz | -13.7 GHz |

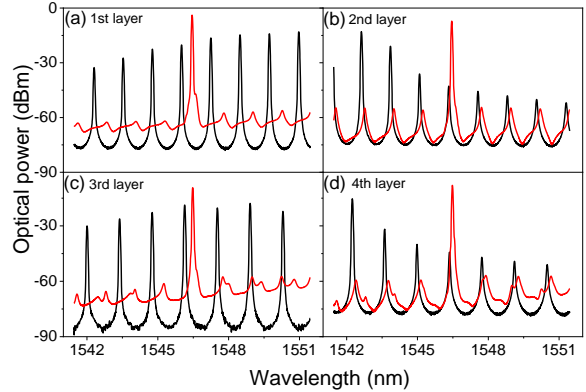

**Fig. 2.** Optical spectra of the four FP lasers with and without optical injection.

The performance of the deep PRC is tested in the real-world task of nonlinear channel equalization in optical fiber communications. The optical signal in optical fibers is distorted by the linear chromatic dispersion and the nonlinear Kerr nonlinearity [36]. The linear distortion is usually mitigated by the feedforward equalizer (FFE) in the digital signal processing (DSP) of the optical receiver [37,38]. On the other hand, the Kerr nonlinearity can be compensated by solving the nonlinear Schrödinger equation [36]. However, common solving algorithms like the digital back propagation are too complex for the DSP implementation [38,39]. An alternative solution is deploying neural networks to compensate the fiber nonlinearity with reduced computational complexity [39-41]. In particular, several literatures have experimentally demonstrated that shallow PRCs were capable to compensate the linear impairments of optical fibers instead of FFEs [33,42-45]. Here we show that the deep PRC has strong ability to mitigate the nonlinear impairments of optical fibers. The nonlinear Schrödinger equation describing the propagation of light in an optical fiber reads [36]:

$$\frac{\partial E}{\partial z} + \frac{\alpha}{2} E + j\frac{\beta_2}{2}\frac{\partial^2 E}{\partial t^2} = j\gamma |E|^2 E \quad (1)$$

where $E(z,t)$ is the slowly varying envelope of the electric field, z is the transmission distance (50 km), α is the attenuation constant (α=0.2 dB/km), $\beta_2$ is the fiber dispersion coefficient (-21.4 $ps^2$/km), and γ is the fiber nonlinearity coefficient (1.2 /(W·km)) [46]. The signal under investigation is a non-return-to-zero (NRZ) signal with a modulation rate of 25 Gbps. The training set consists of 35000 random symbols of {0, 1}, and the testing set consists of 15000 symbols. Each symbol consists of 8 samples, and the tap number of the nonlinear equalizer is set at 21. In the experiment, each measurement is repeated four times, and the mean bit error rate (BER) and the standard deviation are recorded.

Figure 3(a) shows an example of the random NRZ signal sequence sent at the transmitter, with a launch power of 4.0 mW. After a transmission distance of 50 km, nevertheless, the signal received at the receiver in Fig. 3(b) is substantially distorted. Generally, increasing the launch power raises the nonlinear effect, and the signal distortion becomes stronger [36]. The task aims to reproduce the original signal in Fig. 3(a) based on the degraded one in Fig. 3(b). When applying the shallow 1-layer PRC to equalize the received signal in Fig. 3(c), the BER firstly decreases from $5.0\times10^{-3}$ at 1.0 mW down to the minimum value of $3.4\times10^{-3}$ at 100 mW. Above 100 mW, the BER increases with the launch power nonlinearly. Meanwhile, the BERs for launch powers ranging from 80 to 120 mW are below the hard-decision forward error correction (FEC) threshold ($3.8\times10^{-3}$, dashed line) [47]. This is because the PRC inherently owns both linear memory effect and nonlinear memory effect, which are commonly quantified by the linear memory capacity (MC) and nonlinear MC, respectively [24,32]. For low launch powers (see 1.0 mW), the Kerr nonlinearity of the optical fiber is negligible and the signal distortion is mainly induced by the linear chromatic dispersion. Therefore, the impairment compensation only requires the linear memory effect of the PRC, while the nonlinear memory effect plays a negative role. When increasing the launch power (see 20-100 mW), the Kerr nonlinearity appears and hence the nonlinear memory effect of the PRC becomes beneficial to mitigate this nonlinear distortion. The BER reaches the minimum value when the inherent nonlinear memory effect of the PRC matches with the strength of the Kerr nonlinearity of the fiber (see 100 mW). On the other hand, the BER increases when the nonlinear memory capacity is not high enough to compensate the strong fiber nonlinearity (see 120-200 mW). Therefore, the nonlinear equalization ability of the PRC is limited by its maximum nonlinear memory effect. For the deep PRC with two reservoir layers, the BER reduces from $4.4\times10^{-3}$ at 1.0 mW down to the minimum of $1.5\times10^{-3}$ at 120 mW. The BERs for launch powers ranging from 20 to 160 mW are below the FEC threshold. The PRC performance further improves when we increase the PRC depth to three. It is shown that the corresponding BER declines from $4.2\times10^{-3}$ at 1.0 mW down to the minimum of $1.0\times10^{-3}$ at 120 mW. The BERs of the 3-layer PRC are better than those of the 2-layer PRC, for all the studied launch powers ranging from 1.0 mW up to 200 mW. However, the performance of the 4-layer PRC is similar to or slightly worse than that of the 3-layer PRC. This suggests the PRC performance saturates at the depth of three, for this nonlinear signal equalization task. In comparison with the shallow PRC, all the three deep PRCs exhibit better performance for the whole launch power range. In particular, the BERs are significantly reduced in the power range of 80 to 160 mW. Therefore, unlike the shallow PRC, the deep PRCs have very strong ability to mitigate the nonlinearity of optical fibers and hence to improve the transmission signal quality. This compensation ability can be attributed to the strengthened nonlinear memory effect of the deep PRCs, which is discussed in the next section.

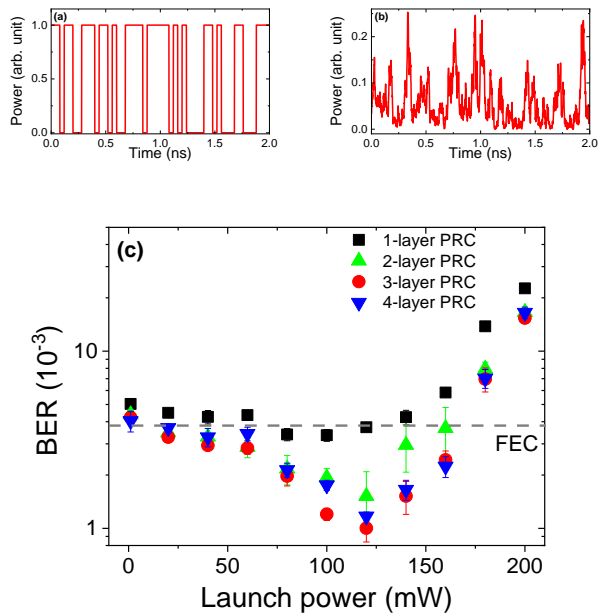

**Fig. 3.** Time sequences of the signal at (a) the transmitter and (b) the receiver. The launch power is 4.0 mW. (c) Performance of the PRCs with different depth. The error bar stands for the standard deviation of the measurement. The dashed line indicates the FEC threshold.

Figure 4 explores the contribution of each reservoir layer in the 3-layer PRC. For each evaluation, only one of the three reservoirs is used for the signal equalization. Therefore, the virtual neuron number becomes 80 instead of 240, both for the training and for the test. It is found that the performance of the second-layer reservoir is generally better than the first-layer one. In particular, the minimum BER of the second-layer reservoir achieved at 120 mW is 2.0×10$^{-3}$. This value further goes down to 1.4×10$^{-3}$ for the third-layer reservoir, which is 2.6 times smaller than the first-layer case (3.7×10$^{-3}$). The different performance of the three hidden layers suggests that the neuron dynamics are different from one layer to another. Generally, the neuron states at the deeper layer are richer than those at the shallower one, which results in the better performance in the former case. This behavior is different to the parallel PRC, where several reservoirs are connected in parallel instead of in series. Our recent experimental work demonstrated that the neuron states in every parallel reservoir were similar to each other [33]. Owing to the rich neuron dynamics in each layer, the nonlinear memory effect in the deep PRC is improved, and thereby the performance of the 3-layer PRC is boosted. In comparison, the FFE commonly used in the DSP of optical receivers only compensates the linear chromatic dispersion of optical fibers [37]. Figure 4 shows that the BER of the FFE increases nonlinearly from 5.2×10$^{-3}$ at 1.0 mW down to the minimum of 3.8×10$^{-2}$ at 200 mW. For the launch power of 120 mW, the BER of the FFE (7.7×10$^{-3}$) is 7.7 times larger than that of the 3-layer PRC (1.0×10$^{-3}$). This comparison proves that the deep PRC can indeed compensate strong nonlinearity of optical fibers. For low launch powers (1 mW), nevertheless, the BER of the 3-layer PRC is only slightly better than that of the FFE. This suggests the deep PRC has similar compensation ability of chromatic dispersion as the FFE.

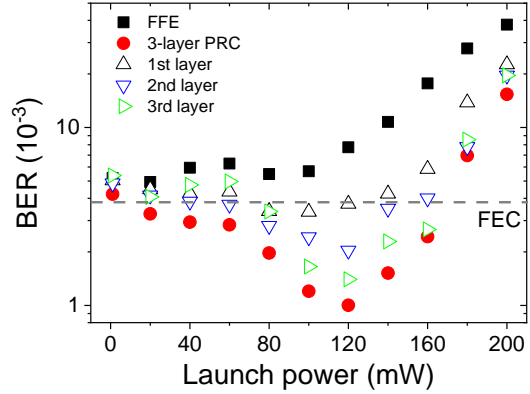

**Fig. 4.** Performance comparison between the 3-layer PRC (dots) and the FFE (squares). The open symbols represent the BERs of each reservoir layer, respectively. The dashed line indicates the FEC threshold.

## 4. DISCUSSION

The experimental results in Fig. 3 and in Fig. 4 have shown that raising the depth of the PRC substantially improves the performance of nonlinearity compensation at high launch powers. However, the deep PRC shows similar performance of linearity compensation as the shallow PRC at low launch powers. In order to understand the behavior, we numerically analyze both the linear MC and the nonlinear MC of the PRC, respectively. The deep PRC model includes four hidden reservoir layers as in the experiment. We assume that the slave lasers in the four layers are all identical to simplify the simulation. The carrier dynamics, the photon dynamics, and the phase of the electrical field are taken into account through the framework of rate equations. Both the optical feedback effect and the optical injection are characterized through the classical Lang-Kobayashi model [48,49]. The main simulation parameters are listed in Table 2. The detailed deep PRC model and other simulation parameters refer to [24]. The linear MC (LMC) measures the ability of the PRC of reproducing the past input signal, which is quantified by [50,51]:

$$MC_L = \sum_{i=1}^{\infty} \frac{\langle u(k-i)y(k)\rangle^2}{\sigma^2[u(k)]\sigma^2[y(k)]} \qquad (2)$$

where the input signal $u(k)$ is a random sequence uniformly distributed in the range of [-1, 1]. $y(k)$ is the corresponding output of the PRC at the step $k$. The aim of the evaluation is to reproduce the input signal $u(k-i)$ shifted $i$-step backward using $y(k)$. $\sigma^2$ represents the variance operation and <> stands for the average operation. On the other hand, the nonlinear MC characterizes the ability of reproducing high-order Legendre polynomials of the input signal, which is defined as:

$$MC_{NL} = \sum_{i=1}^{\infty} \frac{\langle p(k-i) y(k) \rangle^2}{\sigma^2[p(k)]\sigma^2[y(k)]} \quad (3)$$

where the polynomial is $p(k)=[3u^2(k)-1]/2$ for the quadratic MC (QMC), and is $p(k)=[5u^3(k)-3u(k)]/2$ for the cubic MC (CMC), respectively. The aim of the evaluation is to reproduce the polynomial $p(k-i)$ using the PRC output $y(k)$. In addition to LMC, QMC, and CMC, the PRC also has higher-order memory effect and cross memory effect, which are not considered in this work. Figure 5 shows that both the linear MC and the nonlinear MCs rise with the increasing depth of the PRC. The LMC increases from 8.47 for the 1-layer PRC up to 16.87 for the 4-layer PRC. However, the deep PRC in Fig. 3 only slightly reduces the BER at low launch powers. This suggests the LMC of the shallow PRC is already high enough for compensating the chromatic dispersion. On the other hand, the nonlinear QMC increases from 5.25 to 11.27, while the CMC increases from 3.33 to 6.57. The enhanced nonlinear MC can be attributed to the rich neuron states of deep reservoir layers as proved in Fig. 4. As a result, the deep PRC exhibits strong ability in mitigating the nonlinearity of optical fibers. On the other hand, all the three MCs almost saturate as the depth of three, resulting in the performance saturation of the nonlinear signal equalization in Fig. 3.

Table 2. Main parameters of the deep PRC in the simulation

| Parameters | Values |
| --- | --- |
| Laser current | $1.6 \times I_{th}$ |
| Feedback delay | 3.6/2.8/2.0/1.2 ns |
| Feedback ratio | -30 dB |
| Injection ratio | -5 dB |
| Detuning freq. | 0 GHz |
| Neuron per layer | 80 |
| Neuron interval | 10 ps |

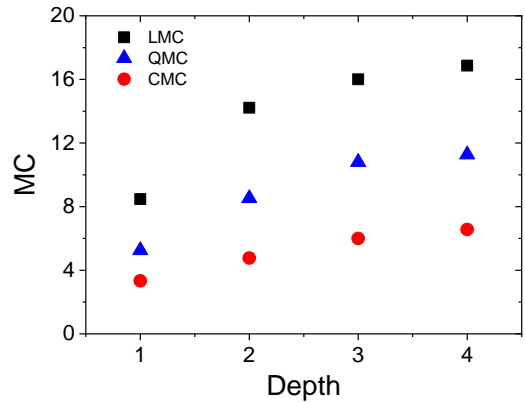

**Fig. 5.** Memory capacity of the PRCs with different depth.

## 5. CONCLUSION

In summary, we have experimentally demonstrated a deep PRC architecture based on the cascading injection-locked Lasers. The connection between successive reservoir layers is all optical, without any OEC or ADC. In addition, this scheme is highly scalable because the laser in each layer provides optical power. The deep PRC with a depth of four is used to solve the real-word problem of nonlinear signal equalization of optical fibers. It is proved that the deep PRC exhibits strong ability to compensate the Kerr nonlinearity of optical fibers and hence to improve the quality of the received signal. In comparison with the linear FFE, the deep PRC reduces the

BER of the transmission link as much as 7.7 times. In comparison with the shallow PRC, the improved performance of the deep PRC is owing to the rich neuron dynamics of the deep reservoir layers, which in turn boosts the nonlinear memory effect. Future work will optimize the operation parameters of the deep PRC, including the injection ratio, the detuning frequency, the feedback ratio, and the feedback delay time.

**Funding**. National Science Foundation of China (NSFC) (61804095). ShanghaiTech University (2022X0203-902-01).

## REFERENCES

1. M. A. Nahmias, T. F. de Lima, A. N. Tait, H. T. Peng, B. J. Shastri, and P. R. Prucnal, "Photonic multiply-accumulate operations for neural networks," IEEE J. Quantum Electron. **26**, 7701518 (2020).
2. Z. Chen and M. Segev, "Highlighting photonics: looking into the next decade," eLight **1**, 2 (2021).
3. C. Huang, V. J. Sorger, M. Miscuglio, M. Al-Qadasi, A. Mukherjee, L. Lampe, M. Nichols, A. N. Tait, T. Ferreira de Lima, B. A. Marquez, J. Wang, L. Chrostowski, M. P. Fok, D. Brunner, S. Fan, S. Shekhar, P. R. Prucnal, and B. J. Shastri, "Prospects and applications of photonic neural networks," Adv. Phys.: X **7**, 1981155 (2022).
4. Y. Shen, N. Harris, S. Skirlo, M. Prabhu, T. Baehr-Jones, M. Hochberg, X. Sun, S. Zhao, H. Larochelle, D. Englund, and M. Soljačić, "Deep learning with coherent nanophotonic circuits," Nat. Photonics **11**, 441-446 (2017).
5. X. Lin, Y. Rivenson, N. Yardimci, M. Veli, Y. Luo, M. Jarrahi, and A. Ozcan, "All-optical machine learning using diffractive deep neural networks," Science **361**, 1004-1008 (2018).
6. X. Xu, M. Tan, B. Corcoran, J. Wu, A. Boes, T. G. Nguyen, S. T. Chu, B. E. Little, D. G. Hicks, R. Morandotti, A. Mitchell, and D. J. Moss, "11 TOPS photonic convolutional accelerator for optical neural networks," Nature **589**, 44–51 (2021).
7. J. Feldmann, N. Youngblood, M. Karpov, H. Gehring, X. Li, M. Stappers, M. Le Gallo, X. Fu, A. Lukashchuk, A. S. Raja, J. Liu, C. D. Wright, A. Sebastian, T. J. Kippenberg, W. H. P. Pernice, and H. Bhaskaran, "Parallel convolutional processing using an integrated photonic tensor core," Nature **589**, 52-58 (2021).
8. E. Goi, X. Chen, Q. Zhang, B. Cumming, S. Schoenhardt, H. Luan, and M. Gu, "Nanoprinted high-neuron-density optical linear perceptrons performing near-infrared inference on a CMOS chip," Light Sci. Appl. **10**, 40 (2021).
9. I. Goodfellow, Y. Bengio, and A. Courville, *Deep learning*. (MIT press, 2016).
10. W. Maass, T. Natschlager, and H. Markram, "Real-time computing without stable states: a new framework for neural computation based on Perturbations," Neural Comput. **14**, 2531-2560 (2002).
11. H. Jaeger and H. Haas, "Harnessing nonlinearity: predicting chaotic systems and saving energy in wireless communication," Science **304**, 78-80 (2004).
12. K. Vandoorne, P. Mechet, T. V. Vaerenbergh, M. Fiers, G. Morthier, D. Verstraeten, B. Schrauwen, J. Dambre, and P. Bienstman, "Experimental demonstration of reservoir computing on a silicon photonics chip," Nat. Commun. **5**, 3541 (2014).
13. M. Nakajima, K. Tanaka, and T. Hashimoto, "Scalable reservoir computing on coherent linear photonic processor," Commun. Phys. **4**, 20 (2021).
14. D. Brunner, M. C. Soriano, C. R. Mirasso, and I. Fuscger, "Parallel photonic information processing at gigabyte per second data rates using transient states," Nat. Commun. **4**, 1364 (2013).
15. J. Moon, W. Ma, J. H. Shin, F. Cai, C. Du, S. H. Lee, and W. D. Lu, "Temporal data classification and forecasting using a memristor-based reservoir computing system," Nat. Electron. **2**, 480-487 (2019).
16. Y. Zhong, J. Tang, X. Li, X. Liang, Z. Liu, Y. Li, Y. Xi, P. Yao, Z. Hao, B. Gao, H. Qian, and H. Wu, "A memristor-based analogue reservoir computing system for real-time and power-efficient signal processing," Nat. Electron. **5,** 672-681 (2022).
17. K. Liu, T. Zhang, B. Dang, L. Bao, L. Xu, C. Cheng, Z. Yang, R. Huang, and Y. Yang, "An optoelectronic synapse based on α-In2Se3 with controllable temporal dynamics for multimode and multiscale reservoir computing," Nat. Electron. **5,** 761-773 (2022).
18. G. Tanaka, T. Yamane, J. B. Héroux, R. Nakane, N. Kanazawa, S. Takeda, H. Numata, D. Nakano, and A. Hirose, "Recent advances in physical reservoir computing: A review," Neural Netw. **115**, 100-123 (2019).
19. C. Gallicchio, A. Micheli, and L. Pedrelli, "Deep reservoir computing: A critical experimental analysis," Neurocomputing **268**, 87-99 (2017).
20. C. Gallicchio, A. Micheli, and L. Pedrelli, "Design of deep echo state networks," Neural Netw. **108**, 33-47 (2018).
21. M. Freiberger, S. Sackesyn, C. Ma, A. Katumba, P. Bienstman, and J. Dambre, "Improving time series recognition and prediction with networks and ensembles of passive photonic reservoirs," IEEE J. Sel. Topics Quantum Electron. **26,** 7700611 (2020).

22. H. Hasegawa, K. Kanno, and A. Uchida, "Parallel and deep reservoir computing using semiconductor lasers with optical feedback," Nanophotonics **12,** 869-881 (2023).

23. M. Goldmann, F. Koster, K. Lüdge, and S. Yanchuk, "Deep time-delay reservoir computing: Dynamics and memory capacity," Chaos **30,** 093124 (2020).

24. B. D. Lin, Y. W. Shen, J. Y. Tang, J. Yu, X. He, and C. Wang, "Deep time-delay reservoir computing with cascading injection-locked lasers," IEEE J. Sel. Top. Quantum Electron. **29,** 7600408 (2023).

25. B. Penkovsky, X. Porte, M. Jacquot, L. Larger, and D. Brunner, "Coupled nonlinear delay systems as deep convolutional neural networks," Phys. Rev. Lett. **123,** 054101 (2019).

26. M. Nakajima, K. Inoue, K. Tanaka, Y. Kuniyoshi, T. Hashimoto, and K. Nakajima, "Physical deep learning with biologically inspired training method: gradient-free approach for physical hardware," Nat. Commun. **13,** 7847 (2022).

27. A. Lupo, E. Picco, M. Zajnulina, and S. Massar, "Fully analog photonic deep Reservoir Computer based on frequency multiplexing," arXiv arXiv:2305.08892 (2023).

28. J. Ohtsubo, *Semiconductor Lasers: Stability, Instability and Chaos* (Springer, 2017).

29. J. Y. Tang, B. D. Lin, Y. W. Shen, R. Q. Li, J. Yu, X. He, and C. Wang, "Asynchronous photonic time-delay reservoir computing," Opt. Express **31,** 2456-2466 (2023).

30. L. Appeltant, M. C. Soriano, G. Van der Sande, J. Danckaert, S. Massar, J. Dambre, B. Schrauwen, C. R. Mirasso, and I. Fischer, "Information processing using a single dynamical node as complex system," Nat. Commun. **2,** 468 (2011).

31. Y. Deng, Z. F. Fan, B. B. Zhao, X. G. Wang, S. Zhao, J. Wu, F. Grillot, and C. Wang, "Mid-infrared hyperchaos of interband cascade lasers," Light Sci. Appl. **11,** 7 (2022).

32. D. Brunner, M. C. Soriano, and G. Van der Sande, *Photonic Reservoir Computing: Optical Recurrent Neural Networks*. (De Gruyter, 2019).

33. R. Q. Li, Y. W. Shen, B. D. Lin, J. Yu, X. He, and C. Wang, "Scalable wavelength-multiplexing photonic reservoir computing," APL Mach. Learn. **1,** 036105 (2023).

34. T. Hülser, F. Köster, L. Jaurigue, and K. Lüdge, "Role of delay-times in delay-based photonic reservoir computing," Opt. Mater. Express **12**, 1214-1231 (2022).

35. C. Wang, K. Schires, M. Osiński, P. J. Poole, and F. Grillot, "Thermally insensitive determination of the linewidth broadening factor in nanostructured semiconductor lasers using optical injection locking," Sci. Rep. **6,** 27825 (2016).

36. G. P. Agrawal, *Nonlinear Fiber Optics* (Springer, 2000).

37. L. Huang, Y. Xu, W. Jiang, L. Xue, W. Hu, and L. Yi, "Performance and complexity analysis of conventional and deep learning equalizers for the high-speed IMDD PON," J. Lightwave Technol. **40**, 4528-4538 (2022).

38. P. J. Freire, Y. Osadchuk, B. Spinnler, A. Napoli, W. Schairer, N. Costa, J. E. Prilepsky, and S. K. Turitsyn, "Performance versus complexity study of neural network equalizers in coherent optical systems," J. Lightwave Technol. **39**, 6085-6096 (2021).

39. Q. Fan, G. Zhou, T. Gui, and A. P. T. Lau, "Advancing theoretical understanding and practical performance of signal processing for nonlinear optical communications through machine learning," Nat. Commun. **11,** 3694 (2020).

40. S. Zhang, F. Yaman, K. Nakamura, T. Inoue, V. Kamalov, L. Jovanovski, V. Vusirikala, E. Mateo, Y. Inada, and T. Wang, "Field and lab experimental demonstration of nonlinear impairment compensation using neural networks," Nat. Commun. **10,** 3033 (2019).

41. C. Huang, S. Fujisawa, T. F. De lima, A. N. Tait, E. C. Blow, Y. Tian, S. Bilodeau, A. Jha, F. Yaman, H. T. Peng, H. G. Batshon, B. J. Shastri, Y. Inada, T. Wang, and P. R. Prucnal, "A silicon photonic–electronic neural network for fibre nonlinearity compensation," Nat. Electron. **4,** 837-844 (2021).

42. A. Argyris, J. Bueno, and I. Fischer, "Photonic machine learning implementation for signal recovery in optical communications," Sci. Rep. **8,** 8487 (2018).

43. J. Vatin, D. Rontani, and M. Sciamanna, "Experimental realization of dual task processing with a photonic reservoir computer," APL Photon. **5,** 086105 (2020).

44. S. Ranzini, R. Dischler, F. Da Ros, H. Bülow, and D. Zibar, "Experimental investigation of optoelectronic receiver with reservoir computing in short reach optical fiber communications," J. Lightwave Technol. **39**, 2460-2467 (2021).

45. I. Estebanez, S. Li, J. Schwind, I. Fischer, S. Pachnicke, and A. Argyris, "56 GBaud PAM-4 100 km transmission system with photonic processing schemes," J. Lightwave Technol. **40,** 55-62 (2022).

46. K. Hammani, B. Wetzel, B. Kibler, J. Fatome, C. Finot, G. Millot, N. Akhmediev, and J. M. Dudley, "Spectral dynamics of modulation instability described using Akhmediev breather theory," Opt. Lett. **36**, 2140-2142 (2011).

47. Meihua Bi, Jiasheng Yu, Xin Miao, Longsheng Li, and Weisheng Hu, "Machine learning classifier based on FE-KNN enabled high-capacity PAM-4 and NRZ transmission with 10-G class optics," Opt. Express **27,** 25802-25813 (2019).

48. R. Lang and K. Kobayashi, "External optical feedback effects on semiconductor injection laser properties," IEEE J. Quantum Electron. **16,** 347-355 (1980).

49. R. Lang, "Injection locking properties of a semiconductor laser," IEEE J. Quantum Electron. **18,** 976-983 (1982).

50. J. Dambre, D. Verstraeten, B. Schrauwen, and S. Massar, "Information processing capacity of dynamical systems," Sci. Rep. **2,** 514 (2012).

51. M. Inubushi, K. Yoshimura, "Reservoir computing beyond memory-nonlinearity trade-off," Sci. Rep. **7,** 10199 (2017).